\documentclass[12pt]{article}

\usepackage{mathrsfs}
\usepackage{graphicx}
\usepackage{color}

\newcommand{\scri}{\mathscr{I}}

\begin{document}

\begin{center}

{\large CLASSICS ILLUSTRATED:}

\

\

{\Large LIMITS OF SPACETIMES}

\vspace{15mm}

{\large Ingemar Bengtsson}

\

{\large S\"oren Holst}

\

{\large Emma Jakobsson}

\vspace{1cm}

{\sl Stockholms Universitet, AlbaNova\\
Fysikum\\
S-106 91 Stockholm, Sweden}

\vspace{15mm}

{\bf Abstract:}

\end{center}

\

\noindent We carefully study the $e \rightarrow m$ and $e \rightarrow 0$ limits of the 
Reissner-Nordstr\"om spacetime using Geroch's definition of limits of spacetimes. This 
is implemented by embedding the one-parameter family of spacetimes in anti-de Sitter 
space, and as a result we obtain metrically correct Penrose diagrams. For 
$e \rightarrow m$ two distinct limits are studied.

\newpage

{\bf 1. Introduction}

\vspace{4mm}

\noindent All students of relativity are familiar with the Reissner-Nordstr\"om 
black hole. Its metric is usually given as 

\begin{equation} ds^2 = - F(r) dt^2 + 
\frac{dr^2}{F(r)} + r^2(d\theta^2 + \sin^2{\theta}d\phi^2) \ , \label{ett} \end{equation}

\begin{equation} F(r) = 1 - \frac{2m}{r} + \frac{e^2}{r^2} \ , 
\hspace{8mm} 0 < e^2 < m^2 \ . \end{equation} 

\noindent In the limit $e \rightarrow 0$ this becomes the Schwarzschild metric, 
and in the limit $e \rightarrow m$ it becomes the metric of an extreme black hole. 
In both cases large regions of spacetime suddenly disappear. Where did they go? 

Even more confusingly, if a simple coordinate transformation is performed first, 
the metric in the limit 
$e \rightarrow m$ is not a black hole metric at all. It is the Bertotti-Robinson 
metric, describing a direct product of 1+1 dimensional anti-de Sitter space with a 
sphere of constant radius. This geometry also appears in the near-horizon 
limit of the extremal black hole. So can the extreme black hole  
be completely bypassed? And are we not told that coordinate changes cannot 
affect anything? 

The student asking such questions is always referred to a classic paper by Geroch 
\cite{Geroch}. 
His paper does indeed explain the matter but leaves many details as exercises for 
the reader. His resolution hinges on one of the core lessons of relativity, namely 
that there does not exist any canonical way to set up a one-to-one correspondence 
between the points of the manifolds underlying two different solutions of the 
field equations. In a situation like the one we are facing (where we want 
to compare the members of a one-parameter family of spacetimes) the best one can 
do is to select one point from each member of the family, erect orthonormal 
frames there, and then regard these points and these frames as ``the same''. 
Only once this has been done can we begin to ask questions about whether a 
given region in one of these spacetimes has a counterpart in some limit---and 
the answers will not be forthcoming unless we are willing to solve for the 
geodesics emerging from that point. 

We will give the answers in a new way. The idea is to embed the relevant 
spacetimes as surfaces in a single embedding spacetime, make sure that 
they touch tangentially at some chosen point, and then watch how they change 
as we vary the parameter $e/m$. Three dimensions suffice for the embedding 
spacetime because it is enough to understand the 1+1 dimensional spacetimes 
coordinatized by $t$ and $r$. So we will be able to actually see what goes on. 
(In principle an exactly analogous treatment can be made for the Kerr 
solution. The results would be similar in many respects, but we would no 
longer be able to understand them just by looking at a picture because 
three dimensions no longer suffice for the embedding.)

A few words on embeddings before we begin. Embeddings of spherically symmetric 
black hole spacetimes in higher dimensional flat spaces have a long history \cite{Ivor}, 
but as far as we know the first actual picture (of the Schwarzschild case) 
was drawn by Marolf \cite{Marolf}. Embeddings into flat reference spacetimes 
have been used as tools to investigate various questions of physical interest \cite{Deser}. 
Perhaps the one closest in spirit to what we do here is a definition of quasi-local 
mass, which uses such an embedding to lay the ghost of diffeomorphism invariance 
to rest \cite{Wang}. 
A global embedding was used to elucidate the nature of the Schwarzschild 
solution \cite{Fronsdal}, but despite some recent progress it is not so easy to 
embed the Reissner-Nordstr\"om spacetime globally in flat space \cite{Paston}. We 
take a different route here, and choose anti-de Sitter 
space as our embedding space. There are two reasons for this. First, it 
enables us to find an embedding that includes a region large enough 
for our purposes. Second, in this way we can see what goes on ``at infinity'' 
where much of the action is. 
In effect our pictures share the advantages of Penrose's conformal diagrams, 
but they contain all the metric information as well.  
An obvious drawback is that it takes a certain 
amount of practice to understand the pictures we draw, but really it is 
quite a small amount, and we give all the necessary detail in the Appendix.
 
\vspace{10mm}

{\bf 2. Coordinate calculations}

\vspace{4mm}

\noindent The maximal analytic extension of the Reissner-Nordstr\"om 
spacetime contains an infinite number of regions of three types. We cover 
one region of each type if we use Eddington-Finkelstein coordinates, in 
which case the metric is 

\begin{equation} ds^2 = - F(r)dv^2 + 2dvdr + 
r^2(d\theta^2 + \sin^2{\theta}d\phi^2) \ , \end{equation}

\begin{equation} F(r) = || \partial_v||^2 = \frac{(r-r_+)(r-r_-)}{r^2} = 
1 - \frac{2m}{r^2} + \frac{e^2}{r^2} \ . \end{equation}

\noindent Here $r$ is the area radius of the round 2-spheres and $v$ labels 
ingoing null geodesics. The roots of the function $F(r)$ are expressed in 
terms of the mass $m$ and the electric charge $e$ by  

\begin{equation} r_\pm = m(1 \pm \epsilon ) \ , \hspace{8mm} 
\epsilon = \frac{\sqrt{m^2-e^2}}{m} \ . \label{epsilon} \end{equation}

\noindent The coordinate system covers a region which is divided into 
three blocks by Killing horizons associated to the Killing vector field 
$\partial_v$. These blocks are 

\begin{equation} \mbox{I}: \ r > r_+ \ , \hspace{6mm} \mbox{II}: \ r_+ > r > r_- \ , 
\hspace{6mm} \mbox{III}: \ r_- > r > 0 \ . \label{regions} \end{equation}

\noindent Using a suitable tetrad the only non-vanishing curvature spinors are  

\begin{equation} \Phi_{11'} = \frac{e^2}{2r^2} \ , \hspace{8mm} 
\Psi_2 = \frac{e^2-mr}{r^4} \ . \label{Weyl} \end{equation} 

\noindent The first of these determines the traceless Ricci tensor, the second 
the Weyl tensor. The function $F(r)$ has a minimum at $r = e^2/m$, corresponding to 
a spacelike hypersurface in block II singled out 
by the fact that $\Psi_2$---and hence the Weyl tensor---vanishes there. 
 
In the limit $\epsilon \rightarrow 1$ ($e \rightarrow 0$) block III 
simply ``disappears''. Moreover a singularity develops in block 
II, so a part of this region disappears as well. If instead we take the 
limit $\epsilon \rightarrow 0$ ($e \rightarrow m$) we obtain an extreme 
black hole with a degenerate Killing horizon at $r = m$. Then block II disappears 
altogether, while the other two blocks ``survive''. 

But we are free to perform the coordinate change 

\begin{equation} r = m + m\epsilon x \ , \hspace{8mm} v = \frac{m}{\epsilon}u 
\ . \label{coordRN} \end{equation}

\noindent Then the Reissner-Nordstr\"om metric becomes  

\begin{equation} ds^2 = m^2\left[ - \frac{(x^2-1)}{(1+\epsilon x)^2}du^2 + 2dudx 
+ (1+\epsilon x)^2(d\theta^2 + \sin^2{\theta}d\phi^2)\right] \ . \end{equation}

\noindent If we now take the limit $\epsilon \rightarrow 0$ ($e \rightarrow m$) 
we obtain the metric 

\begin{equation} ds^2 = m^2\left[ - (x^2-1)du^2 + 2dudx 
+ d\theta^2 + \sin^2{\theta}d\phi^2\right] \ . \end{equation}

\noindent In this limit $\Psi_2 = 0$, so the metric has become 
conformally flat \cite{Carroll}. It describes a patch of the Bertotti-Robinson 
spacetime, a direct product of 1+1 dimensional anti-de Sitter space with 
a sphere of constant radius. All three blocks contribute in this limit.

The story gets an extra twist if we first go to the 
limit $e =m$ in the original coordinates, then perform the coordinate 
transformation (\ref{coordRN}) but this time with a dimensionless parameter 
$\epsilon$ which is not defined by eq. (\ref{epsilon}), and finally let 
this parameter go to zero. The end result is known as the near-horizon 
geometry of the extremal black hole. Its metric is 

\begin{equation} ds^2 = m^2\left[ - x^2du^2 + 2dudx + d\theta^2 + 
\sin^2{\theta}d\phi^2\right] \ . \end{equation}

\noindent In fact this is the Bertotti-Robinson metric once again, but 
in a coordinate system covering a different region than that we had 
above.

Exactly parallel calculations can be made for the Kerr spacetime \cite{Helgi}. 

\vspace{10mm}

{\bf 3. Geroch's explanation}

\vspace{4mm}

\noindent Evidently the coordinate calculations are rather confusing, so 
we turn to Geroch's paper for clarification. In his setup a one-parameter 
family of spacetimes is really a five dimensional manifold foliated by 
the members of the family. The question whether a limit exists is the question whether 
it is possible to add a boundary to this larger manifold. The reason why 
a given one-parameter family may admit several different limits can be traced 
back to the fact that there exists no natural identification point-by-point 
of two different spacetimes. Any such identification must be added as an 
extra piece of structure (a vector field) to the five dimensional manifold. 
Geroch then proves an important rigidity theorem. It says that in order to 
make the ambiguities go away, we must single out one fiducial point in each 
spacetime and regard them as (so to speak) the same point, and moreover we 
must single out one orthonormal tetrad at each such point and regard 
them as the same tetrad at the same point. The question 
whether any given point has a counterpart in the limit then turns into the task of 
finding a broken geodesic connecting the given point to the fiducial point, 
and studying its behaviour in the limit. This is a well defined procedure 
because a geodesic from the fiducial point is uniquely determined in 
terms of the structure we have given, and the tetrad can be parallelly transported 
along it to determine the remaining segments of the broken geodesic. 

\begin{figure}
	\begin{center}
	\leavevmode
	\includegraphics[width=50mm]{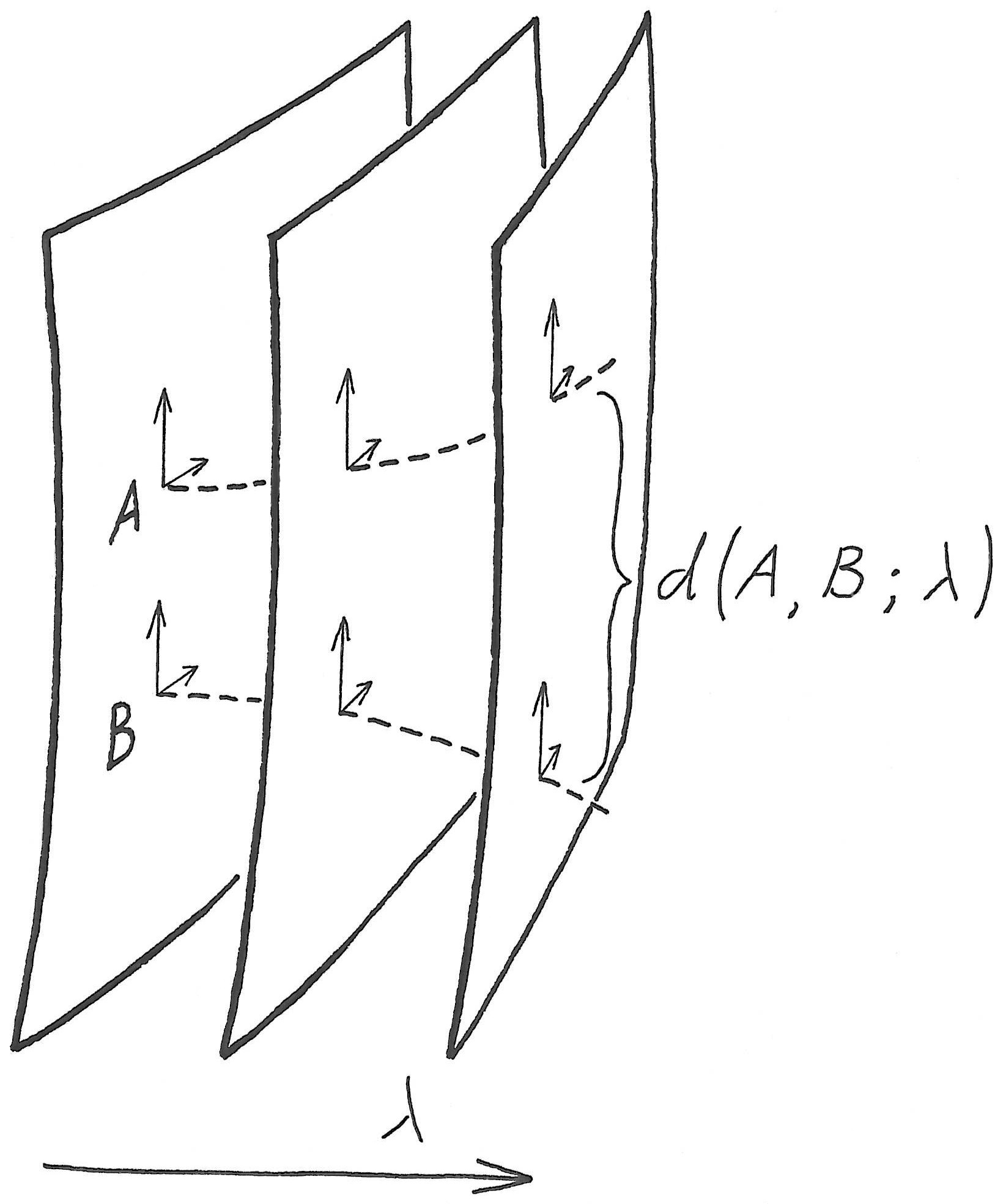}
	\end{center}
	\caption{\small Geroch's five dimensional manifold is foliated by a 
        family of spacetimes parametrized by a single parameter $\lambda$. 
        We can select points, such as $A$ and $B$, 
on the leaves and regard them as the same on each leaf. But if the geodesic 
distance between $A$ and $B$ diverges in the limit they cannot both have 
counterparts in the same limiting spacetime.}
	\label{fig:limitsFig1}
\end{figure}

It is now easy to see how different choices of fiducial points can lead 
to different limits. 
A glance at Fig. \ref{fig:limitsFig1} should be enough.

\begin{figure}[ht]
	\begin{center}
	\leavevmode
	\includegraphics[width=55mm]{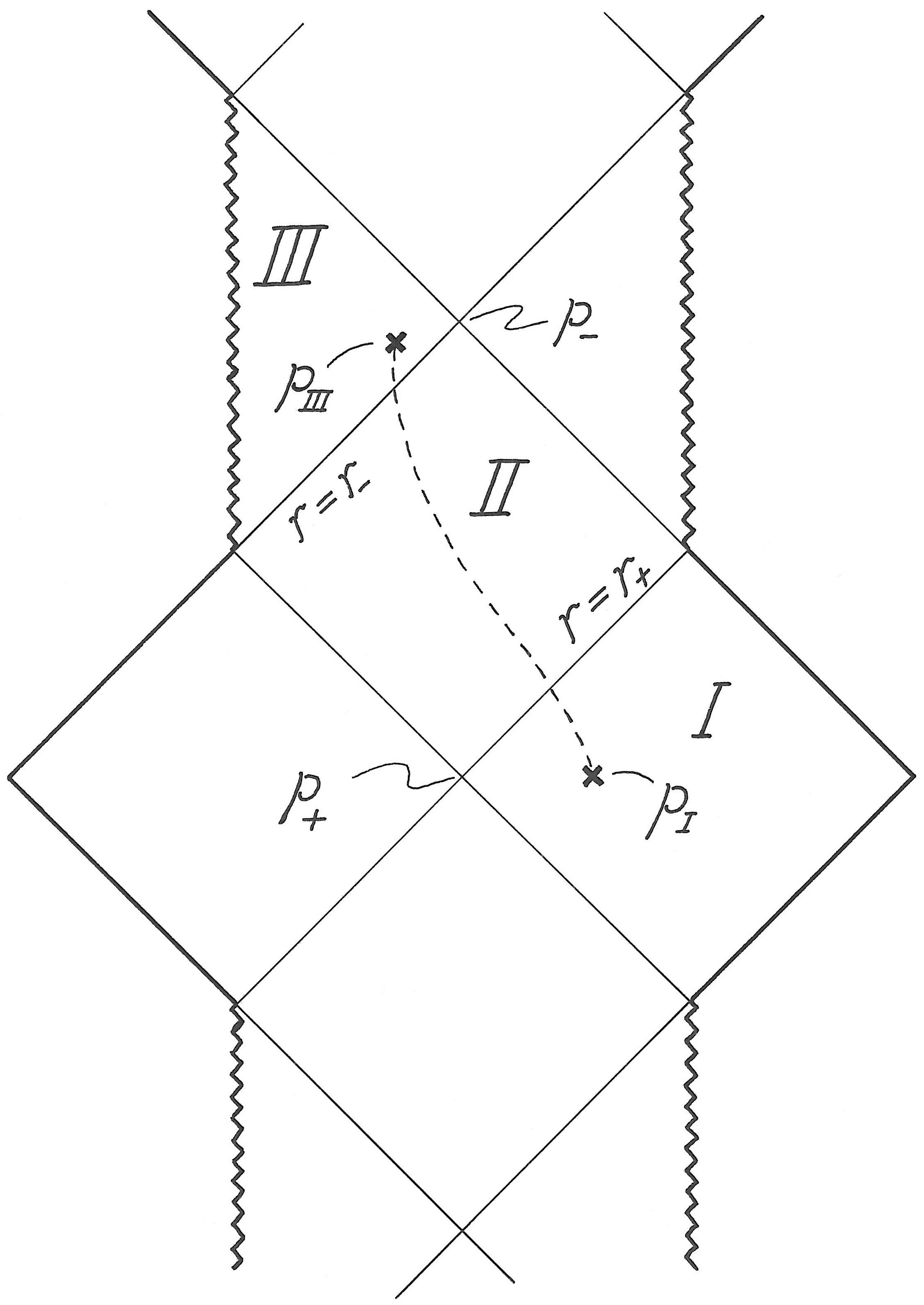}
	\end{center}
	\caption{\small A Penrose diagram of the Reissner-Nordstr\"om spacetime. There is 
an event horizon at $r = r_+$ and a Cauchy horizon at $r = r_-$. They have 
bifurcation points $P_\pm$. We have also chosen a point $P_{I}$ in the exterior 
region I and a point $P_{III}$ in the interior region III, since they will 
play a role in the discussion.}
	\label{fig:limitsFig2}
\end{figure}

\frenchspacing
In the case at hand this difficulty does in fact appear. Consider the selection 
of points in the Reissner-Nordstr\"om spacetime that appear in Fig. 2. \nonfrenchspacing We can 
draw geodesics between them (for which purpose the detailed discussion of radial 
Reissner-Nordstr\"om geodesics given by Brigman is helpful \cite{Brigman}). The 
first observation then is that 

\begin{equation} \lim_{e\rightarrow m} d(P_+,P_I) = \infty \ . \end{equation}

\noindent This means that at most one of these points can survive in this limit. 
Similarly

\begin{equation} \lim_{e\rightarrow m} d(P_-,P_{III}) = \infty \ . \end{equation}

\noindent On the other hand 

\begin{equation} d(P_+,P_-) = \pi m \ , \end{equation}

\noindent independent of $e$. So we expect that if $P_+$ appears in the limit, so 
does $P_-$. This cannot 
be quite true though, because the point $P_-$ does in fact disappear in the 
$e \rightarrow 0$ limit. We will return to this issue below. A  
geodesic between $P_I$ and $P_{III}$ is naturally divided into three segments, 
and we are particularly interested in the segment that crosses region II: 

\begin{equation} d(P_I,P_{III}) = d(P_I,r_+) + d(r_+,r_-) + d(r_-,P_{III}) \ . \end{equation}

\noindent For the middle segment we find 

\begin{equation} \lim_{e \rightarrow m} d(r_+,r_-) = 0 \ . \end{equation}

\noindent The two points on the horizons will either coincide in the limit, or become 
null separated. In the former case region II will be absent. If the latter is 
true something dramatic must happen in the limit to the pair of timelike separated 
points $P_I$ and $P_{III}$---as will become clear later. 

Following this line of argument through we recover the two different 
limits that we encountered in the coordinate calculation. See Fig. 3. Complete 
understanding requires 
further calculations. Thus Geroch raises the question how close to the Cauchy 
horizon a point in region II may come, and still have a counterpart in the 
Schwarzschild limit $e \rightarrow 0$. 

\begin{figure}[ht]
	\begin{center}
	\leavevmode
	\includegraphics[width=0.8\textwidth]{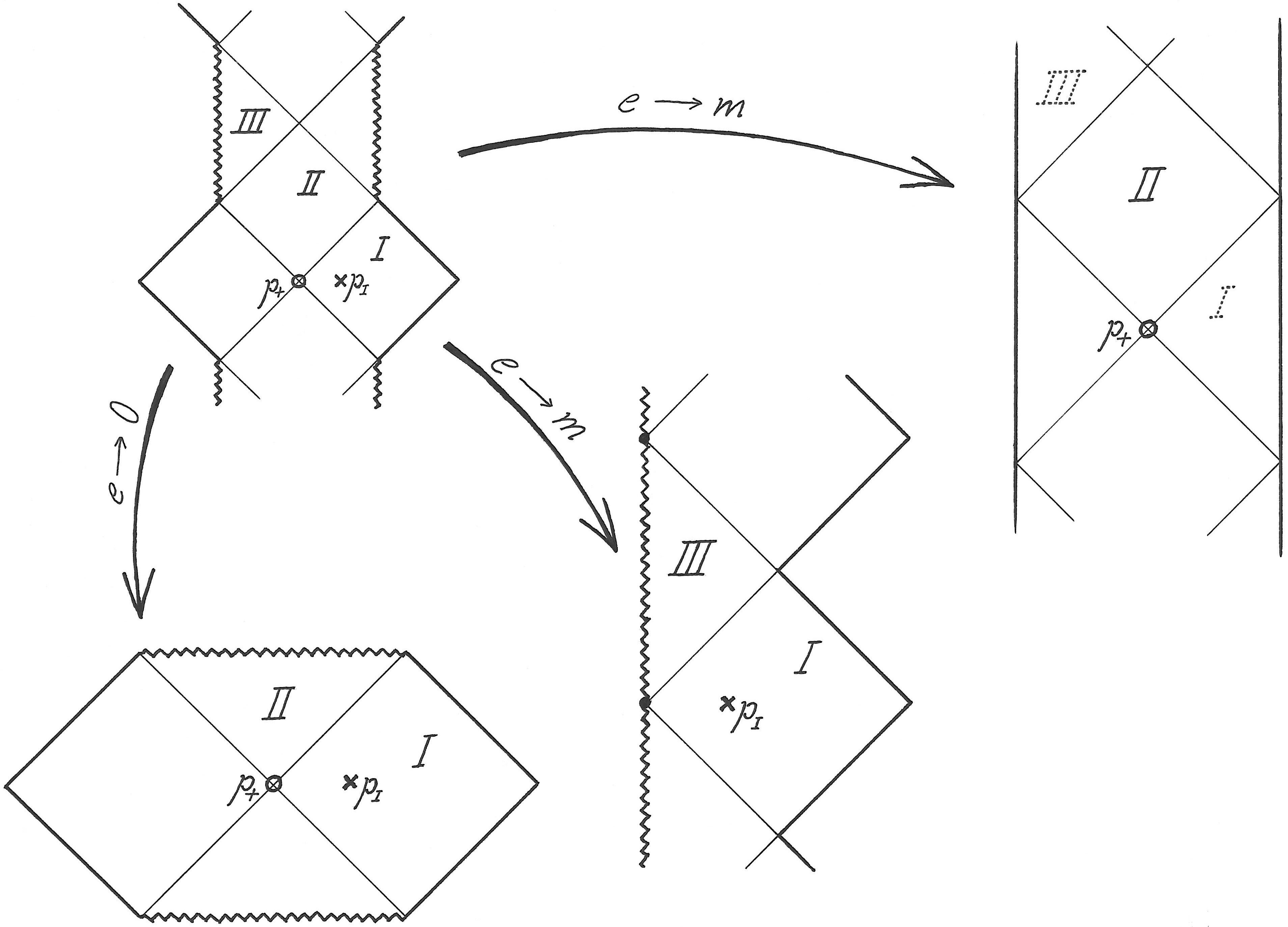}
	\end{center}
	\caption{\small Penrose diagrams of three limiting spacetimes. For the first, 
        $e \rightarrow 0$ and either $P_+$ or $P_I$ is the preferred point. This is 
        the Schwarzschild spacetime. Region III and a part of region II have 
        disappeared. For the second, $e \rightarrow m$ and 
$P_I$ is the preferred point. This is an extreme black hole. Region II has 
disappeared. For the third, $e \rightarrow m$ and $P_+$ is the preferred point. 
This is the Bertotti-Robinson spacetime, anti-de Sitter times a sphere. Although 
there is a region I and a region III in this limit, they are the ghosts of 
departed quantities in the sense that they come from the near-horizon 
regions in the original spacetime.}
	\label{fig:limitsFig3}
\end{figure}

The main purpose of the rest of our paper is to make sure that we 
really {\it see} what goes on---seeing as opposed to just giving nodding agreement 
to the formulas. Once this has been done it will be easy to answer Geroch's 
question, and we do so in section 6.

\vspace{10mm}

{\bf 4. An embedding into anti-de Sitter space}

\vspace{4mm}

\noindent Our way of implementing Geroch's procedure is to embed all spacetimes 
into a fixed ambient space. All members of the one-parameter family are to touch 
at a definite point in this reference space, and moreover their tangent planes are 
to coincide there. In this way all the requirements for Geroch's rigidity theorem 
to hold will be met. 

For our purposes it will be enough to embed the 1+1 dimensional spacetimes obtained 
by ignoring their spherical parts. We will refer to them as black hole surfaces. 
For our ambient spacetime we choose 2+1 dimensional anti-de Sitter space, 
trusting that the reader familiarizes herself with the Appendix in order to 
be able to read the resulting pictures. We choose, quite arbitrarily, the 
curvature scale $l^2 = 1$ in the ambient space, and as a result our pictures will 
depend (inessentially but noticeably) on the parameter $m$ that determines 
the intrinsic spacetime 
geometry. We will in fact set $m = 1$ when drawing the pictures. This represents 
no loss of generality since we can make the pictures independent of the value of 
$m$ by adjusting the scale $l^2$, although we will not do so explicitly.  

For anti-de Sitter space we use the embedding coordinates presented in the 
Appendix. They are called that because anti-de Sitter space itself is presented 
as embedded in a flat space of one dimension higher. In the first embedding 
we place the bifurcation point $P_+$ at the point $(X,Y,U,V) = (0,0,0,1)$ for 
all values of $e$. We refer to this point in anti-de Sitter space as 
its ``origin'', because it serves as the origin of the sausage coordinate system 
(again we refer to the Appendix). 
 
Explicitly we perform the embedding 
of the black hole surface by setting 

\begin{equation} Y^2 - U^2 = a^2F(r) \ , \hspace{8mm} X^2-V^2 = - 1 - a^2F(r) \ . 
\end{equation}

\noindent Here $a$ is a dimensionful constant which will eventually be set equal to $1/\kappa$, where $\kappa$ is the surface gravity of the event horizon. 
We complete the description by 

\begin{equation} (Y,U) = \left\{ \begin{array}{cll} a\sqrt{F(r)}(\cosh{\frac{\tau}{a}}, \sinh{\frac{\tau}{a}}) & \mbox{if} & 
F(r) > 0 \\ \\ a\sqrt{-F(r)}(\sinh{\frac{\tau}{a}},\cosh{\frac{\tau}{a}}) & \mbox{if} & F(r) < 0 
\end{array} \right.  \label{Y} \end{equation}

%\begin{equation} U = \left\{ \begin{array}{cll} a\sqrt{V(r)}\sinh{\frac{\tau}{a}} & \mbox{if} & 
%V(r) > 0 \\ \\ a\sqrt{-V(r)}\cosh{\frac{\tau}{a}} & \mbox{if} & V(r) < 0 
%\end{array} \right. \label{U} \end{equation}

\begin{equation} (X,V) = \left\{ \begin{array}{cll} \sqrt{1+a^2F(r)}(\sinh{g},\cosh{g}) & \mbox{if} & 
1 + a^2F(r) > 0 \ \ \\ \\ \sqrt{-1-a^2F(r)}(\cosh{g}, \sinh{g}) & \mbox{if} & 1 + a^2F(r) < 0 \ \ 
. \end{array} \right. \label{X} \end{equation}

%\begin{equation} V = \left\{ \begin{array}{cll} \sqrt{l^2+a^2V(r)}\cosh{g} & \mbox{if} & 
%1 + a^2V(r) > 0 \ \ \\ \\ \sqrt{-1-a^2V(r)}\sinh{g} & \mbox{if} & 1 + a^2V(r) < 0 \ .
%\end{array} \right. \end{equation}

\noindent The function $g = g(r)$ is so far undetermined. 
Some signs still need care since in fact we do not want to 
restrict ourselves to positive $Y$ when $F(r) > 0$ (for instance). 
By construction $\tau$ is a coordinate along the flow lines of the Killing vector 
$J_{YU}$. Thus we have identified the static Killing vector field on the black hole 
surface with the Killing vector $J_{YU}$ in anti-de Sitter space. In this way 
our construction gains some rigidity. 

We still have to choose the form of $g(r)$. We insist that $g(r_+) = 0$. With 
the above definitions (and a prime 
to denote derivatives with respect to $r$) the induced metric becomes 

\begin{eqnarray} ds^2 = dX^2 + dY^2 - dU^2 - dV^2 = \hspace{45mm} \nonumber \\ \ \\  
= - Fd\tau^2 + \frac{dr^2}{F}\left( 
F(1+a^2F)g^{\prime 2} + \frac{\frac{a^2F^{\prime 2}}{4}}{1 
+ a^2F}\right) \ . \nonumber \end{eqnarray}

\noindent To make this coincide with the metric on the black hole surface---see eq. 
(\ref{ett})---we must impose the differential equation   

\begin{equation} g^{\prime 2} = 
\frac{1 + a^2F - \frac{a^2F^{\prime 2}}{4}}{F(1+a^2F)^2} \ . \label{gprimad} \end{equation} 

\noindent If the right hand side is negative the embedding fails. Since the denominator 
contains the factor $F(r)$ it changes sign at the horizons, so there is a potential 
problem with this. We address it by choosing \cite{Marolf}

\begin{equation} \frac{1}{a} = \kappa = \frac{1}{2}\frac{dF}{dr}_{|r = r_+} = 
\frac{r_+-r_-}{2r_+^2}\ . \end{equation}

\noindent Thus $\kappa$ is the surface gravity of the outer horizon. This solves the 
problem because the factor $(r-r_+)$ cancels out, and the embedding equation becomes 

\begin{equation} g^{\prime 2} =  \frac{\kappa^2}{4r_+^4}\frac{(r_+-r_-)^2(r^2+r_+^2)(r+r_+)r^2 
+ 4r_+^4(r^4-r_+r_-)(r-r_-)}{(r-r_-)((\kappa^2+1)r^2-(r_++r_-)r + r_+r_-)^2} \ . 
\label{emb2} \end{equation}

\noindent On the right hand side we have a fifth order polynomial divided by 
another (easily factorized) fifth order polynomial. 
We see immediately that the denominator changes sign at the inner horizon $r = r_-$. Hence we cannot cover both the outer and inner horizons of the black hole surface 
in this manner. There are two additional zeroes of the denominator between 
the horizons leading to a divergent but positive expression. This is allowed.  

The sign of the numerator at the outer horizon $r = r_+$ is easily 
derived. It is found to be non-negative there whenever 

\begin{equation} r_+ -r_- +r_+^3 - r_- \geq 0 \ . \end{equation}

\noindent In particular it is non-negative whenever $r_+^2 \geq 1$, and this can 
be arranged for all values of $e^2/m^2 \leq 1$ by choosing $m^2 \geq 1$. (We 
remind the reader that the length scale of anti-de Sitter 
space has already been set equal to one.) If $m$ is so small that $r_+^3$ can be 
ignored in comparison to the linear terms the embedding condition at $r = r_+$ implies 

\begin{equation} r_+ \geq 2r_- \hspace{5mm} \Leftrightarrow \hspace{5mm} 
\frac{e^2}{m^2} \leq \frac{8}{9} \ . \end{equation} 

\noindent Because of our choice of a length scale $m << 1$ corresponds to a black 
hole surface embedded in a nearly flat spacetime, and we conclude---correctly 
\cite{Marolf2}---that a similar embedding into Minkowski space fails when 
$e^2/m^2 > 8/9$. Embedding into anti-de 
Sitter space is therefore a distinct improvement.

If one really wants to see what happens at the bifurcation point on the inner 
horizon one can perform another embedding there, but we will not do so here. Our 
task is to solve the differential equation for $g$, see what the embedded 
surface looks like, and address the question of limits that is our stated aim. 

\vspace{10mm}

{\bf 5. The Schwarzschild surface}

\vspace{4mm}

\noindent Before attending to the details it is helpful to observe that we have 
arranged the embedding so that the Killing field $J_{YU}$ is everywhere tangential 
to the surface. In effect then it is enough to determine what the embedding looks 
like in the Poincar\'e disk at $t = 0$ and in the 1+1 dimensional anti-de Sitter 
section at $y = 0$. (We are slipping into the language of sausage coordinates, 
which we spell out in the Appendix.) The embedded surface will be swept out 
from these two curves by the Killing flow lines. The event horizon is a Killing 
horizon, and will appear as two null geodesics emerging from the origin of anti-de 
Sitter space. 

\begin{figure}[ht]
	\begin{center}
	\leavevmode
	\includegraphics[width=55mm]{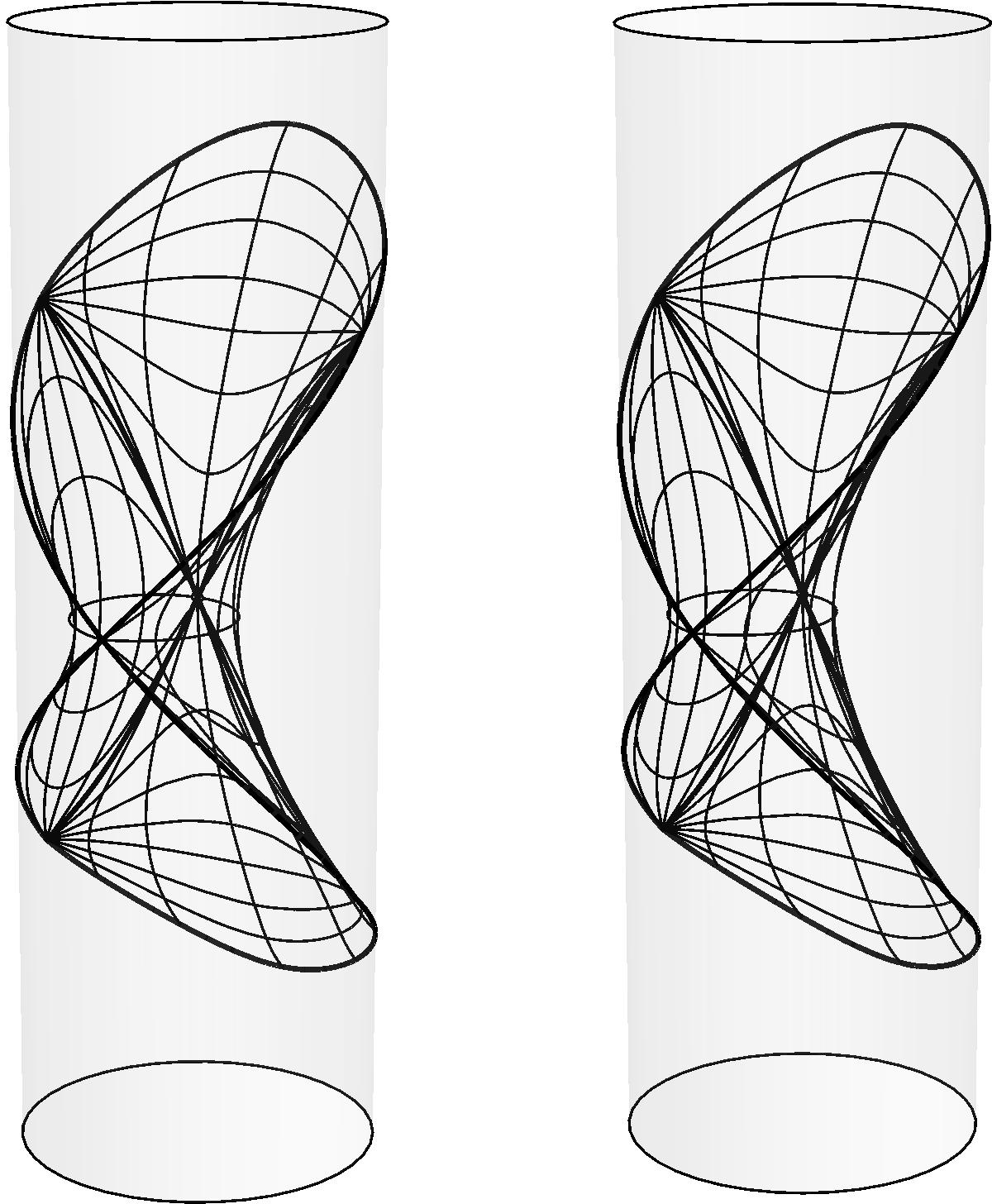}
	\end{center}
	\caption{\small Here is the Schwarzschild surface as a stereogram. The 
        three dimensional picture emerges, in the middle, after appropriate focussing 
        on the part of the reader. The singularity forms a spacelike curve on the 
        anti-de Sitter boundary, and Schwarzschild infinity meets this boundary 
        in a pair of null curves intersecting at $i^0$. The event 
        horizon forms two intersecting null geodesics inside the cylinder. For a 
        step by step construction of this surface see Fig. \ref{fig:skivor}.}
	\label{fig:limitsFig4}
\end{figure}

We begin with the Schwarzschild spacetime ($e =0$), as an exercise. We solve 
the differential equation (\ref{emb2}) using Mathematica, plot the intersection of the 
embedded black hole surface on the Poincar\'e disk $t = 0$ and on 
the timelike plane $y = 0$, and finally draw the full three-dimensional 
picture. Of course the result will depend on our choice of the parameter 
$m$. On the whole the various pictures that we want to draw look nicer 
if we choose $m = 1$, so we settle for this value and show the results 
in Fig. \ref{fig:limitsFig4}. 

The picture looks just as one would expect, given that one knows the embedding 
into flat Minkowski space \cite{Marolf}. A second look is worthwhile though---what 
we see is a metrically correct Penrose diagram. Its conformal boundaries lie on the 
anti-de Sitter $\scri$ as null lines. In the picture they happen to meet at 
$i^0$, which is not unexpected. The singularity appears as a 
spacelike curve on $\scri$ . At first sight this may seem surprising, since timelike 
Schwarzschild geodesics can reach it in finite time. However, these curves are 
not even close to being geodesics in anti-de Sitter space. Indeed it is possible 
to reach infinity along timelike curves in anti-de Sitter space, and to do 
so in finite time, provided that one is able to spend rocket fuel at a suitably increasing rate. The same is true in Minkowski space \cite{Geroch2}.   

\vspace{10mm}

{\bf 6. The first two limits of the Reissner-Nordstr\"om surface}

\vspace{4mm}

\begin{figure}[ht]
	\begin{center}
	\leavevmode
	\includegraphics[width=55mm]{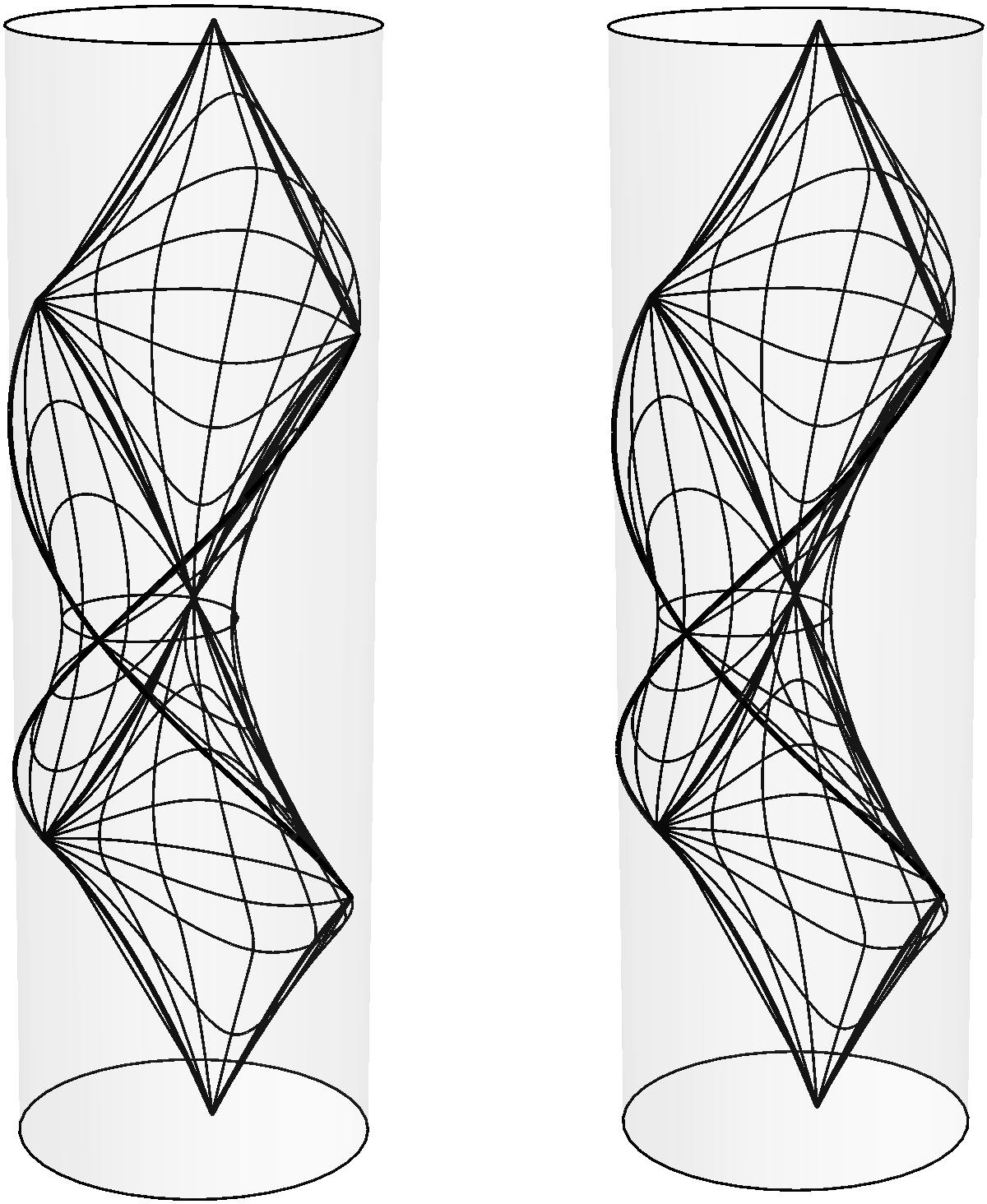}
	\end{center}
	\caption{\small Here is the Reissner-Nordstr\"om surface, again as a stereogram, 
        with four blocks of the Penrose diagram included. The embedding fails at the 
        inner horizon, which is seen as two pairs of null geodesics at the top and bottom 
        of the surface. The value of $e/m$ has been set equal to 0.9. Compare to 
        Fig. \ref{fig:skivor}.}
	\label{fig:limitsFig5}
\end{figure}

\begin{figure}[ht]
	\begin{center}
	\leavevmode
	\includegraphics[width=110mm]{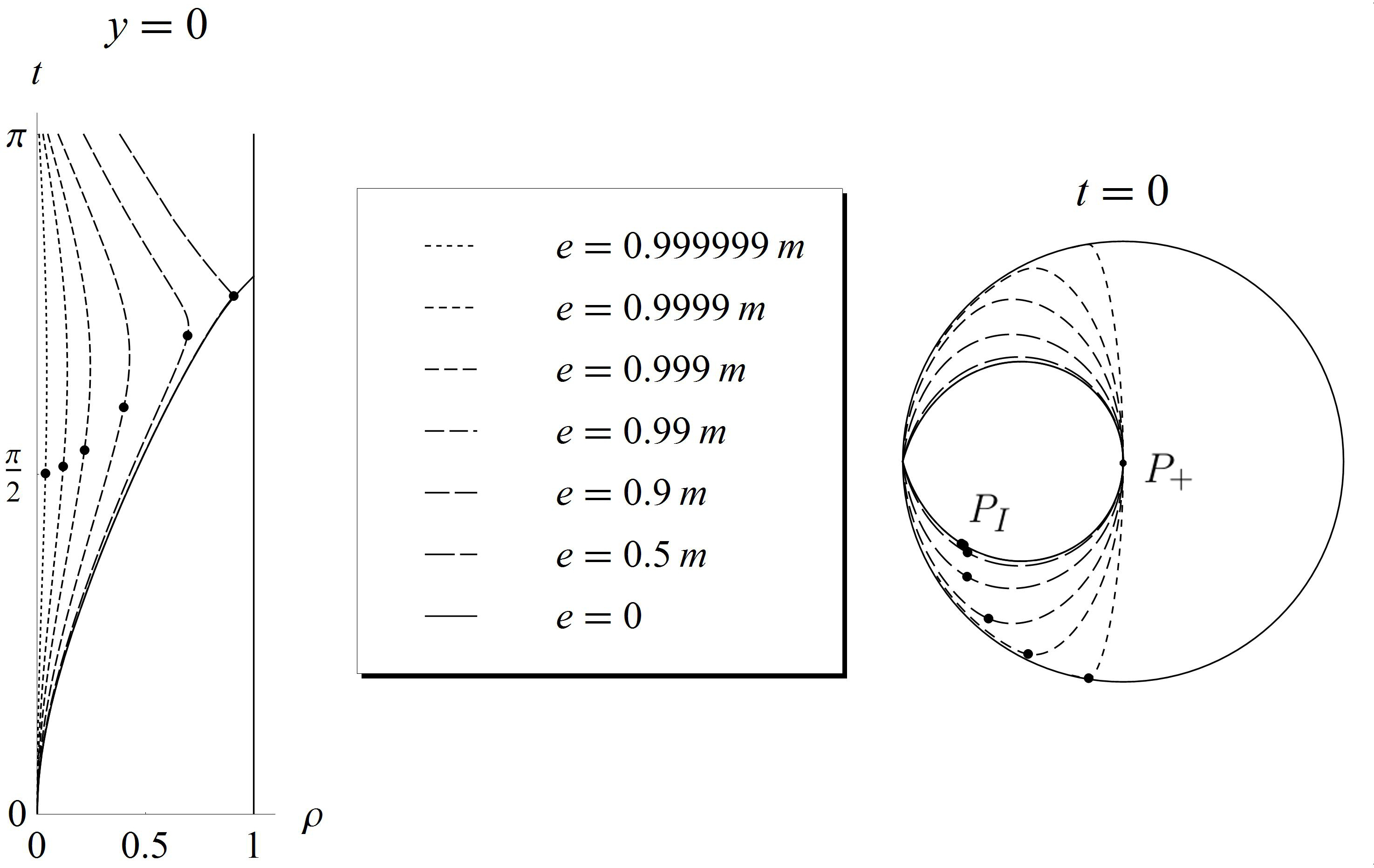}
	\end{center}
	\caption{\small The three dimensional pictures can be understood by looking 
        at the intersection of the surface with the timelike plane at $y = 0$ (left; only 
        its upper half is shown) and the 
        disk at $t = 0$ (right). Various values of $e/m$ are included. In the plane we have 
        placed a dot at $r = e^2/m$ to see where this point is heading as $e \rightarrow 
        0$. The bend is very sharp, but still smooth, already for $e = 0.5m$. 
        In the disk we have marked the point $P_I$ (see the text) 
        to see where it goes (namely to infinity) 
        as $e \rightarrow m$. All curves are tangential at the anti-de Sitter origin.}
	\label{fig:skivor}
\end{figure}

\noindent We now come to the Reissner-Nordstr\"om case. Its black hole surface 
centred at the bifurcation point $P_+$ is shown in Fig. \ref{fig:limitsFig5}. 
However, in order to understand its limits we need to consult 
Fig. \ref{fig:skivor}, which contains the same information but for a variety 
of values of $e/m$. 

Look first at the intersection of the black hole surface with the disk at $t = 0$. 
What we see in Fig. \ref{fig:skivor} is that it approaches the line $x = 0$ as 
$e \rightarrow m$, and then continues to $i^0$ in the immediate vicinity of $\scri$ . 
This means that all points at finite distance from the bifurcation point $P_+$ 
are pushed closer and closer to $\scri$. In the picture this is illustrated for a 
point $P_I$ chosen to sit at that value of $r$ for which 
$r^2/F(r)$ has a minimum, which is also that special value of $r$ for which the four 
dimensional geometry admits a null geodesic at constant $r$. This point sits at a 
finite distance from the event horizon, but quite close to it in astronomical terms. 
In the (singular) limit all points 
originally at finite spacelike distance from $P_+$ disappear. Looking next 
at the intersection with the timelike plane $y = 0$ we see that it comes closer 
and closer to the vertical as $e/m$ increases. In the limit it becomes the vertical. 
The black hole surface itself tends to the 1+1 dimensional anti-de Sitter space at 
$x = 0$. 
In this way it becomes clear that the Reissner-Nordstr\"om spacetime does indeed 
approach the Bertotti-Robinson spacetime in the limit $e \rightarrow m$, given 
that the bifurcation point $P_+$ is included in the entire family. 

The limit $e \rightarrow 
0$ is interesting too, and we can now answer Geroch's question concerning what 
points in region II that survive in this limit. In the plane 
$y = 0$ the Reissner-Nordstr\"om surface appears as a curve with a 
bend which will reach $\scri$ in the limit $e \rightarrow 0$, and in the limit 
all points above the bend disappear. The question is then at what $r$-value the 
bend occurs. Referring back to eq. (\ref{X}), and noting that $1 + a^2F < 0$ 
in the interesting region, we see that the bend occurs at 

\begin{equation} X^\prime = g^\prime \sqrt{-1-a^2F(r)}\sinh{g} -  
\frac{a^2F^\prime (r)}{2\sqrt{-1-a^2F(r)}}\cosh{g} = 0 \ . \end{equation}

\noindent Now keep $r = e^2/m$ fixed, so that $F^\prime (r) = 0$. In fact we place 
ourselves at the minimum of the function $F$. We know that 
$F(e^2/m)$ diverges in the limit $e \rightarrow 0$. Referring back to 
eq. (\ref{gprimad}), and noting that $g$ is finite here, we see that 

\begin{equation} \lim_{e\rightarrow 0}X^\prime (e^2/m) = \lim_{e\rightarrow 0} \frac{\sinh{g}}{\sqrt{-F(e^2/m)}} = 0 \ . \end{equation} 

\noindent This locates the position of the bend in the limit, and shows that the 
surface is discontinued precisely at $r = e^2/m$. This then is the answer to the 
question left open by Geroch. 

Incidentally the equation $r = e^2/m$ defines a rather interesting hypersurface 
in the Reissner-Nordstr\"om spacetime, since the four dimensional Weyl tensor 
vanishes there. See eq. (\ref{Weyl}). 

One comment concerning Fig. \ref{fig:skivor} is that we have to go to rather 
high values of $e/m$ in order 
to see any substantial departure from the Schwarzschild case in the vicinity 
of the bifurcation point $P_+$. This is understandable. 
The principal curvatures $k_1$, $k_2$ of the embedded surface, which is what one 
``sees'' in the picture, are related to the intrinsic curvature $R$ of the two 
dimensional black hole surface through the Gauss equation 

\begin{equation} k_1k_2 = 1 + \frac{R}{2} = 1 + \frac{2mr - 3e^2}{r^4} \ . 
\end{equation}

\noindent If we set $r = r_+$ and then expand in $e/m$ we obtain for the 
Reissner-Nordstr\"om surface that 

\begin{equation} k_1k_2 = 1 + \frac{1}{4m^2}\left( 1 - \frac{3}{16}\frac{e^4}{m^4} 
+ \dots \right) \ .  \end{equation}

\noindent We see that the effects are not noticeable until the number $(e/m)^4$ is. 

\vspace{10mm}

{\bf 7. The extreme limit of the Reissner-Nordstr\"om surface}

\vspace{4mm}

\noindent It remains to recover the other $e \rightarrow m$ limit of the 
Reissner-Nordstr\"om black hole, in which the limit spacetime is an extreme 
black hole. To do so we focus our attention on a point $P_I$ inside region I. 
We will arrange the embedding in such a way that this point is placed at 
the origin of anti-de Sitter space. Moreover---in order to ensure that there 
is a special tetrad attached to that point---we have to do the embedding in 
such a way that all black hole surfaces meet tangentially there, for all 
values of $e/m$. This happened almost automatically in our first embedding, 
but will require conscious effort this time. 

The first question is what point to choose. We must make some choice for 
definiteness, even if it matters only in a rather superficial way. Again we settled 
for that value of $r$ where the four dimensional geometry admits a null 
geodesic at constant $r$, namely where $r^2/F(r)$ has a minimum. This point 
(which figures also in the disk shown in Fig. \ref{fig:skivor}) is 
now to be placed at the anti-de Sitter origin, and all the black hole surfaces 
are required to meet tangentially there. Then Geroch's 
requirements for a well defined limit are met.

\begin{figure}[ht]
	\begin{center}
	\leavevmode
	\includegraphics[width=0.8\textwidth]{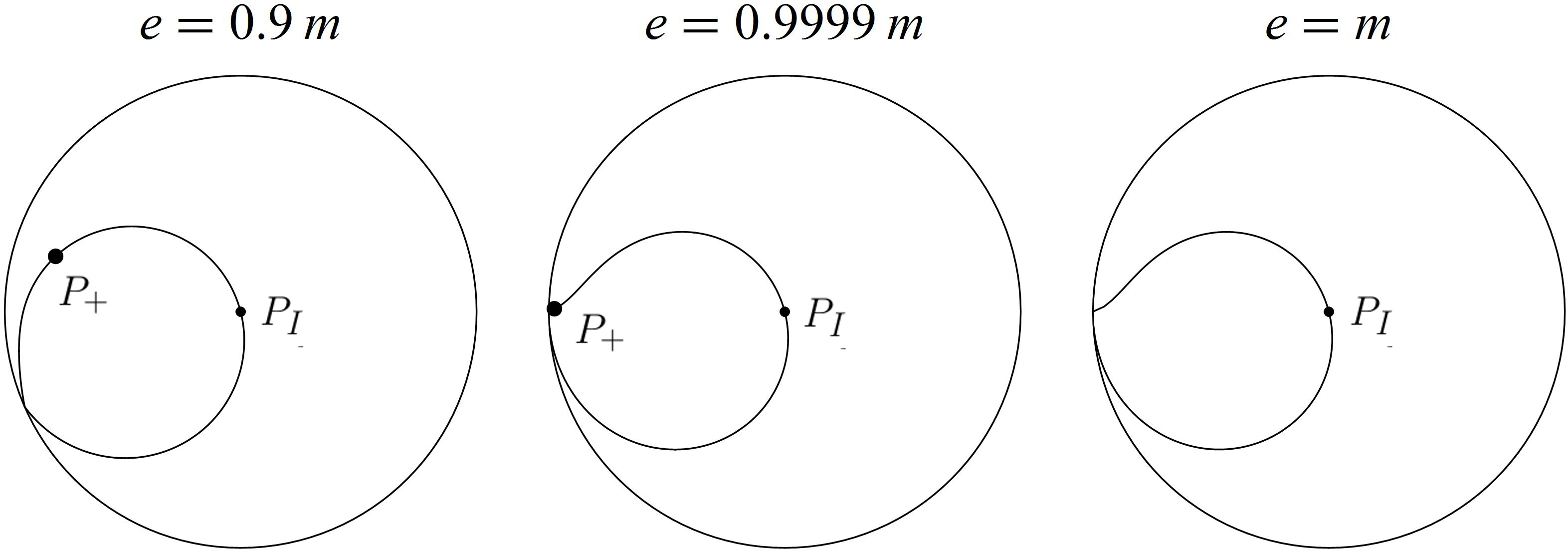}
	\end{center}
	\caption{\small A point $P_I$ in region I is placed at the anti-de 
        Sitter origin, and all the black hole surfaces meet tangentially there.
        When we take the limit $e \rightarrow m$ the bifurcation 
        point moves to infinity and an extreme 
        black hole results.}
	\label{fig:gransskivor}
\end{figure}

The limiting procedure is illustrated in Fig. \ref{fig:gransskivor}. We started with 
a convenient embedding of the extreme black hole surface, and then applied an 
isometry---in the Poincar\'e disk, a M\"obius transformation---to the previous 
embeddings, ensuring that all these black hole surfaces have the point $P_I$ placed 
at the anti-de Sitter origin and that they all meet tangentially there. As in the 
previous case the distance between $P_I$ and $P_+$ diverges in the limit, but 
this time it is the bifurcation point $P_+$ that is pushed off to infinity. 

\begin{figure}
	\centering
	\includegraphics[width=55mm]{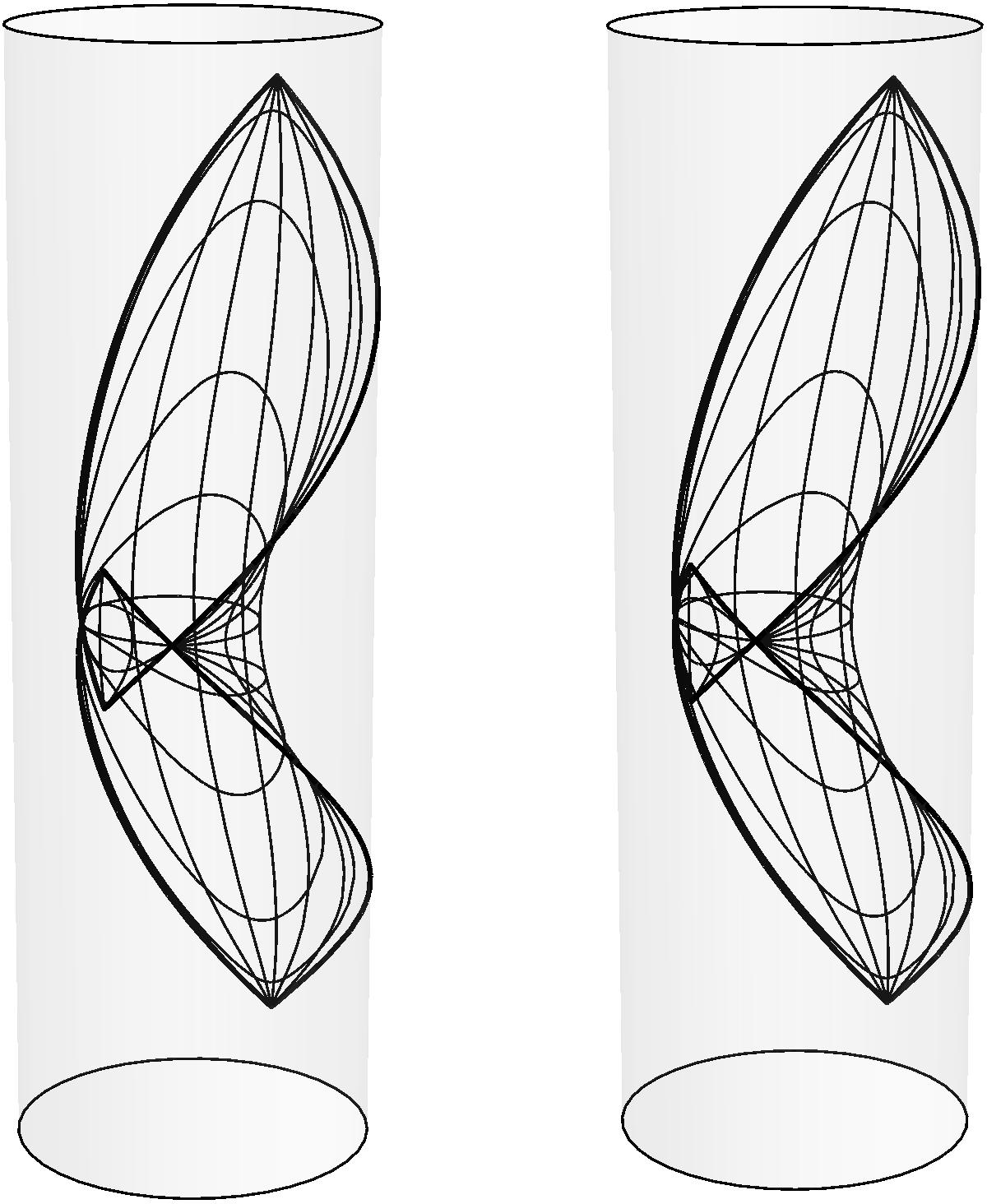}
	\caption{\small A Reissner-Nordstr\"om black hole surface with $e/m = 0.9$ and 
        a point $P_I$ in its exterior placed at the anti-de Sitter origin. Only its 
        exterior regions are shown, so the surface is bounded by the event horizon. 
        Metrically the two exterior regions have the same size, but one of them sits 
        close to the conformal boundary. The vanishing interior region II is not shown, but if it were, it would be seen that it collapses to a null geodesic connecting two points on $\scri$ as $e \to m$.}
	\label{fig:e0.9_skev}
\end{figure}

\begin{figure}
	\centering
	\includegraphics[width=55mm]{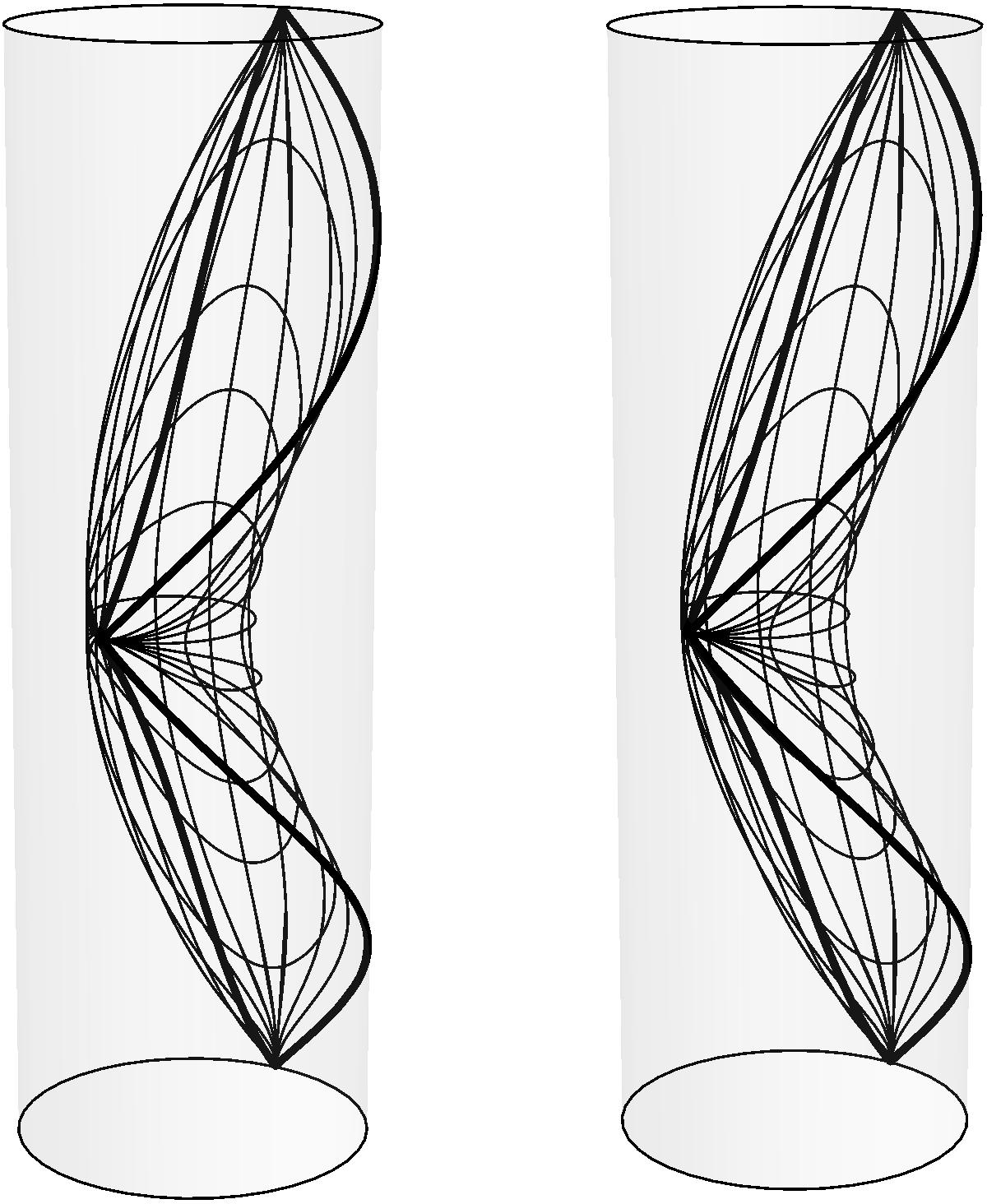}
	\caption{\small The extreme black hole surface. The embedding does not cover 
        its interior and the horizon is a pair of null geodesics in anti-de Sitter space. 
        Only one exterior region survived the limit.}
	\label{fig:extrem}
\end{figure}

\frenchspacing
Stereograms 
of a Reissner-Nordstr\"om black hole arranged in this way, and of the resulting extreme 
black hole surface, are shown in Figs. \ref{fig:e0.9_skev} - \ref{fig:extrem}. \nonfrenchspacing
The pictures look skew because the geometry of the extreme black hole 
is obviously not symmetric under reflection in $P_I$. (It does 
enjoy a discrete conformal symmetry though \cite{CT}). Also its Killing 
vector field is of a different type than that of the non-extreme black hole, 
when considered as a Killing vector field in anti-de Sitter space. In the terminology 
of ref. \cite{Stefan} it is type $II_a$ rather than type $I_b$. 

\vspace{10mm}

{\bf 8. Conclusions}

\vspace{5mm}

\noindent We have shown that embeddings into a fixed ambient spacetime can be 
used to visualize the way a one-parameter family approaches a limiting spacetime. 
In particular this setting makes transparent the nature of the conditions that 
Geroch had to impose in order to make such limits well defined, as well as the reasons 
why a given family of spacetimes may admit several different limits.

Our use of anti-de Sitter space, rather than flat space, as embedding space has 
the advantages that it becomes simpler to perform the embedding, and 
that our pictures can be regarded as metrically correct Penrose diagrams. The 
embedding covers four blocks of the Reissner-Nordstr\"om spacetime, which is 
enough for our purposes.

The reader may have noticed that we have studiously avoided the question whether 
more than two limiting spacetimes exist (when $e \rightarrow m$). We do not know the 
answer, but we do know that the possibilites are tightly constrained. The only 
interesting case where the answer appears to be known is that of the $m \rightarrow \infty$ 
limits of the Schwarzschild spacetime. There a coordinate free calculation 
based on the Cartan-Karlhede scalars has shown that five different limits 
exist \cite{Paiva}. But that method has its drawback too, since it rather obscures 
the physical nature of the limits. It would be interesting to combine methods. 

\

\

\noindent \underline{Acknowledgements}: We thank Helgi R\'unarsson, Istvan Racz, 
Jos\'e Senovilla, and Stefan \AA minneborg for their encouraging interest in limits.

\

\

\newpage

{\bf Appendix: Visualizing anti-de Sitter space}

\vspace{5mm}

\noindent A first remark is that visualizing anti-de Sitter is in many 
ways easier than visualizing Minkowski space, because the conformally 
compactified version is so much simpler to see.

\

\noindent {\it Sausage coordinates:} The standard description of 2+1 dimensional anti-de 
Sitter space is as the hypersurface 

\begin{equation} X^2 + Y^2 - U^2 - V^2 = - l^2 \end{equation}

\noindent embedded in a four dimensional space with the indefinite metric 

\begin{equation} ds^2 = dX^2 + dY^2 - dU^2 - dV^2 \ . \end{equation}

\noindent There is a natural length scale $l$. We set $l = 1$ by fiat. 
The best way to visualize this spacetime is to use the sausage coordinates 
$(t,x,y)$ defined by 

\begin{equation} X = \frac{2x}{1-\rho^2} \hspace{6mm} 
Y = \frac{2y}{1-\rho^2} \hspace{6mm} 
U = \frac{1+\rho^2}{1-\rho^2}\sin{t} \hspace{6mm} 
V = \frac{1+\rho^2}{1-\rho^2}\cos{t} \ , \end{equation} 

\noindent where $\rho^2 = x^2 + y^2 < 1$. The anti-de Sitter metric is then 

\begin{equation} ds^2 = - \left( \frac{1+\rho^2}{1-\rho^2}\right)^2dt^2 + 
\frac{4}{(1-\rho^2)^2}(dx^2 + dy^2) \ . \end{equation}

\noindent The resulting picture is that of a solid 
cylinder, with $\scri$ at its boundary and with spatial slices at constant $t$ 
being copies of the Poincar\'e disk. 

\begin{figure}[t!]
	\begin{center}
	\leavevmode
	\includegraphics[width=0.5\textwidth]{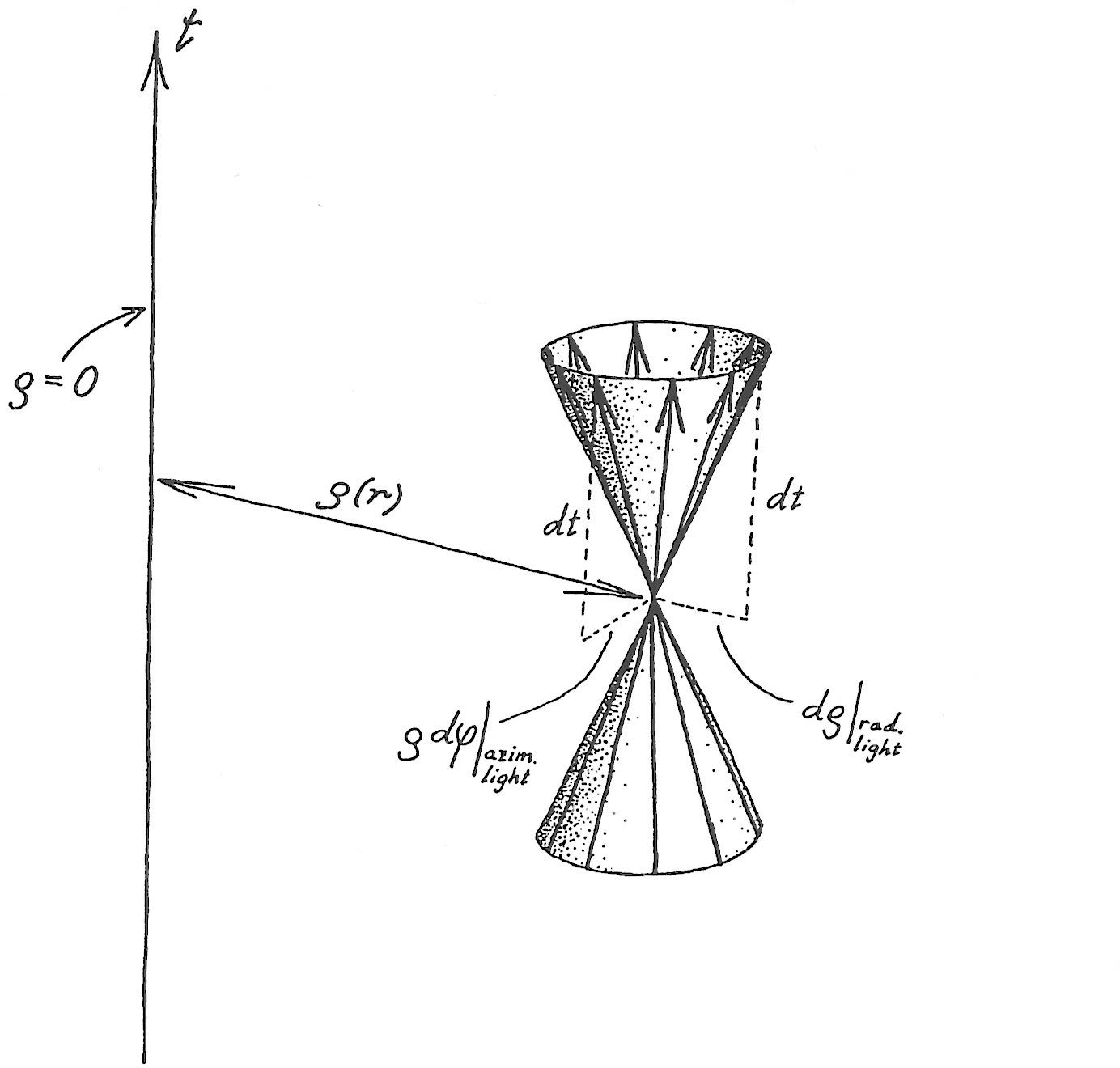}
 \end{center}
	\caption{\small The advantage of sausage coordinates is that the light 
        cone is a circular cone at every point, even if its opening angle depends 
        on the radial coordinate $\rho$.}
	\label{fig:limit1}
\end{figure}

As a rule of thumb, all calculations are simplified by the embedding coordinates 
$X, Y, U, V$, while the sausage coordinates 
$t, x, y$ are superior for drawing pictures.

\

\begin{figure}
	\begin{center}
	\leavevmode
	\includegraphics[width=0.3\textwidth]{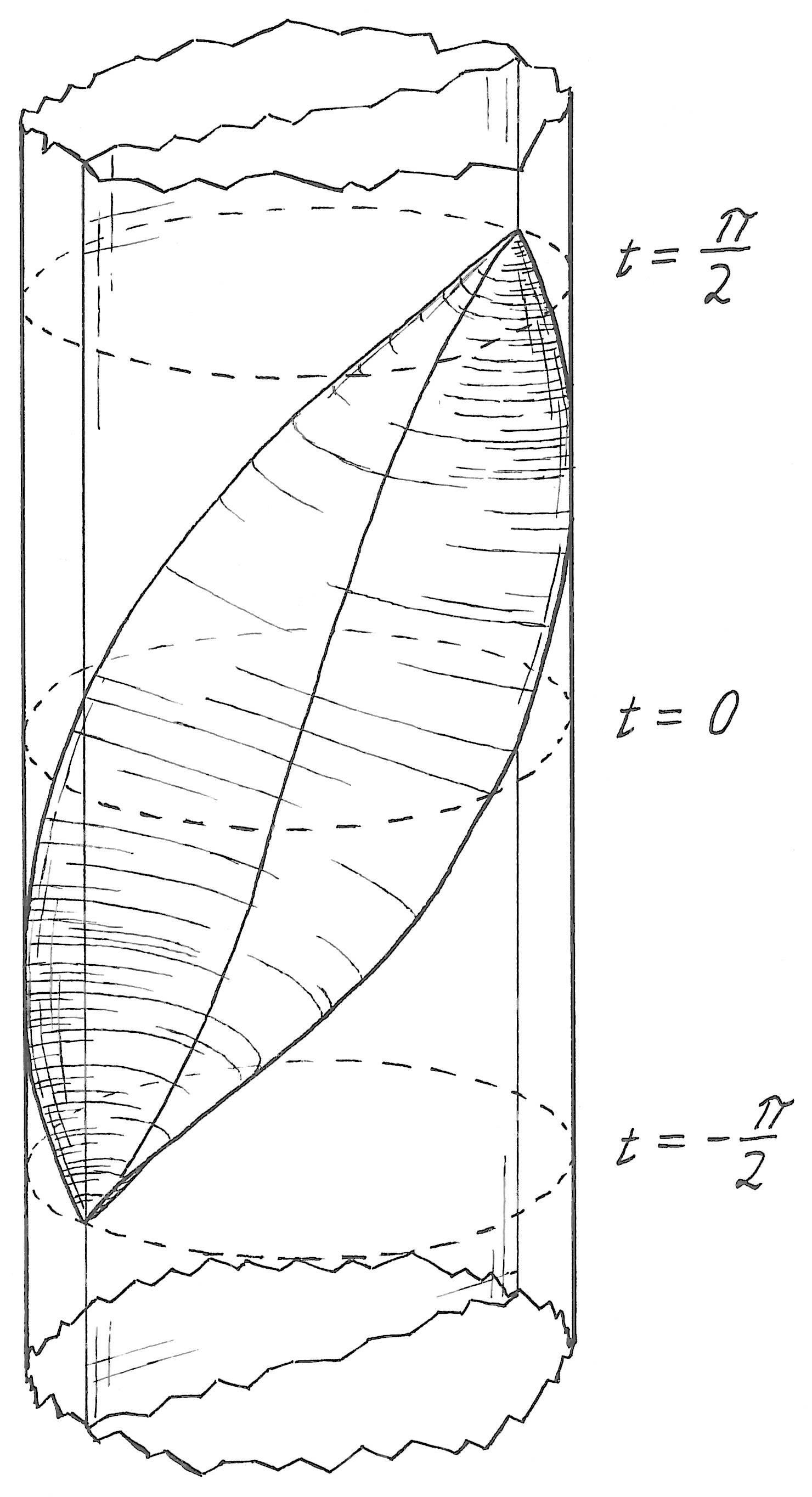}
	\end{center}
	\caption{\small A light cone with its vertex on $\scri$, or, a null plane 
	in anti-de Sitter space. This picture is useful when thinking about the causal structure.}
	\label{limitsFig5}
\end{figure}

\begin{figure}
	\begin{center}
	\leavevmode
	\includegraphics[width=1.0\textwidth]{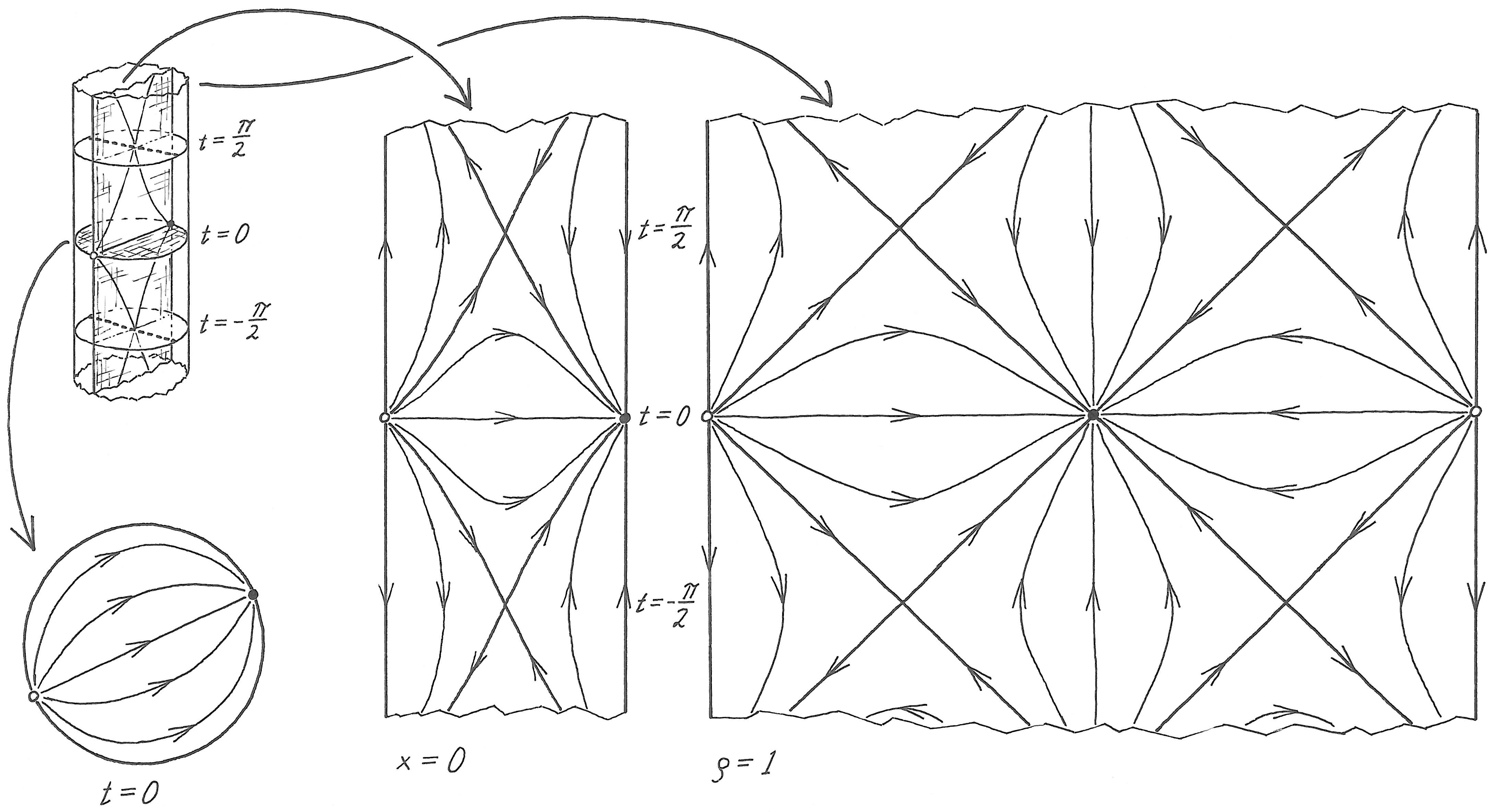}
	\end{center}
	\caption{\small The flow of the Killing vector $J_{YV}$, depicted on the Poincar\'e 
	disk at $t = 0$, on the timelike plane $x = 0$, and on the conformal boundary $\rho = 1$.}
	\label{limitsFig4}
\end{figure}

\noindent {\it The pictures:} It looks like a salami sliced by Poincar\'e 
disks, and we assume 
that the reader is familiar with the geodesics and the isometries in these 
disks. Since space is 
manifestly conformally flat a light cone looks like a right circular cone in 
the neighbourhood of any point, although its opening angle depends on the radial 
coordinate $\rho$ (Fig. \ref{fig:limit1}). On $\scri$ light rays slope 45 degrees. It 
is useful to know what a lightcone with a vertex on $\scri$ looks like, and 
this is shown in 
Fig. \ref{limitsFig5}. The flow of the Killing vector $J_{YV}$ is shown in Fig. \ref{limitsFig4}. 
Its Killing horizon has a bifurcation line at $Y = V = 0$. At $U = 0$ ($t = 0$) it acts 
as a hyperbolic translation in the Poincar\'e disk, and at $V = 0$ ($t = \pi/2$) 
it acts like a Lorentz boost. This information 
should be kept in the back of one's mind as one looks at the illustrations in 
the main text.

\

\noindent {\it Isometries:} 
A good grasp of the isometries is useful, and easily obtained due to the curious 
coincidence that 2 + 1 dimensional anti-de Sitter space is 
identical to the group manifold of $SL(2, {\bf R})$. It can therefore be 
parametrized as 

\begin{equation} g = \left( \begin{array}{cc} U+Y & X+V \\ X-V & U-Y \end{array} 
\right) \ , \hspace{8mm} \det{g} = - X^2 - Y^2 + U^2 + V^2 = 1 \ . \end{equation}

\noindent We placed the identity element of the group at $t = \pi/2$, well separated 
from the Poincar\'e disk at $t = 0$ which in itself is a conjugacy class in the 
group. All isometries are of the form $g \rightarrow g_1gg_2^{-1}$, and the 
isometry group splits into factors according to 

\begin{equation} SO(2,2) = (SL(2,{\bf R})\times SL(2, {\bf R}))/Z_2 \ . \end{equation}

\ 

\noindent {\it The conformal boundary:} The conformal boundary $\scri$ , at $\rho = 1$, 
is a useful platform from which to read the pictures. The group $SO(2,2)$ acts 
as the conformal group there. The two factors into which it splits act independently 
as M\"obius transformations on the null coordinates $v = t+\phi$ and $u = t-\phi$. 
A full description can be found elsewhere \cite{Stefan}. 

\

\begin{figure}[h!]
	\begin{center}
	\leavevmode
	\includegraphics[width=55mm]{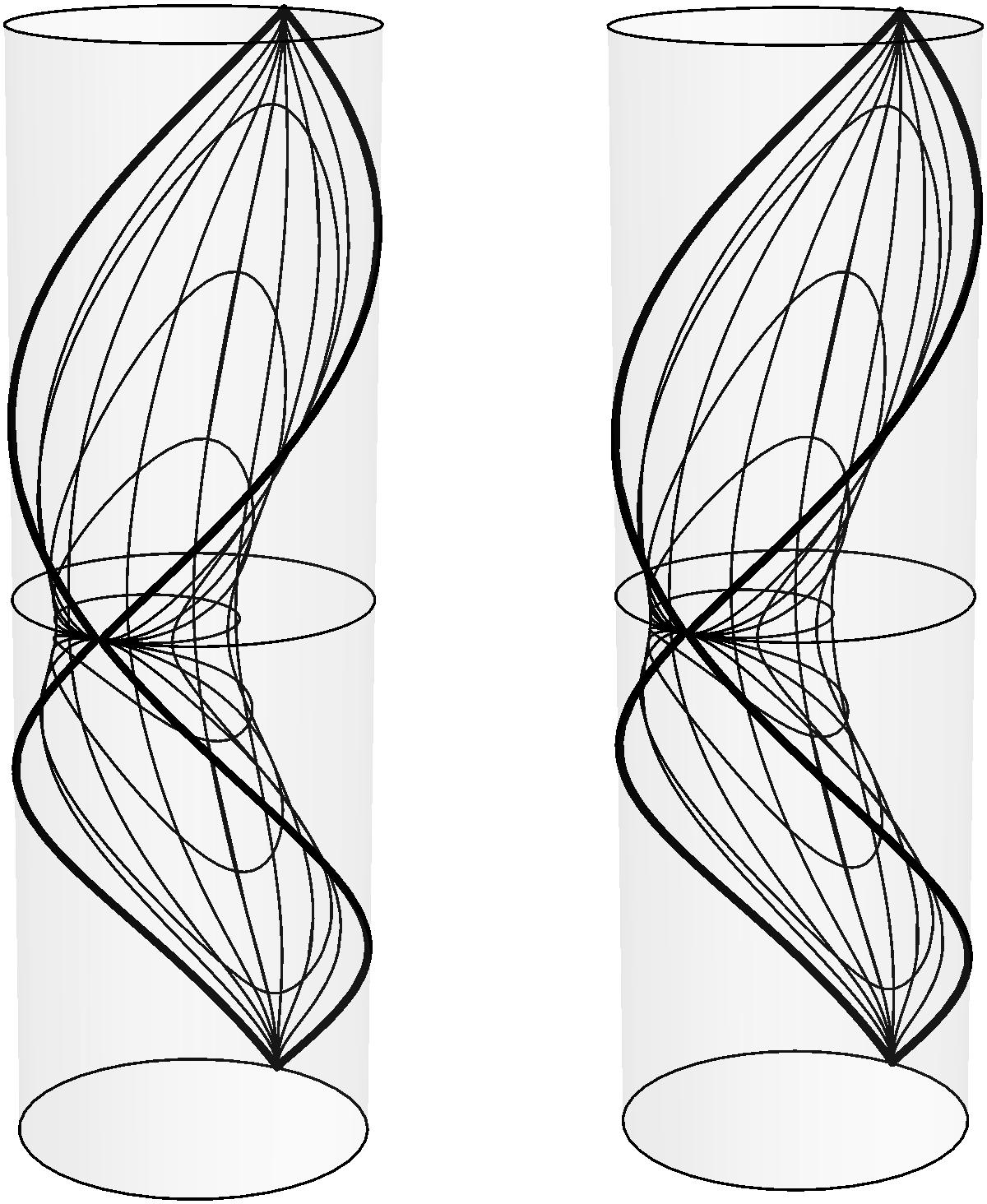}
	\end{center}
	\caption{\small $1 + 1$ dimensional Minkowski space embedded in 
        $2+1$ dimensional anti-de Sitter space, as a stereogram. The 
        Poincar\'e disk at $t = 0$ is shown.}
	\label{fig:Minkowski}
\end{figure}

\noindent {\it A reference surface:} Finally, to give the reader something 
to compare the black hole surfaces to, we give the embedding of Minkowski 
space into anti-de Sitter space in Fig. \ref{fig:Minkowski}. It cuts the 
Poincar\'e disk at $t=0$ in a circle touching the boundary at a point. Such 
a circle is called a horocircle.


\newpage 

{\small
 
\begin{thebibliography}{99}

\bibitem{Geroch} R. Geroch, {\it Limits of spacetimes}, Commun. Math. Phys. {\bf 13} 
(1969) 180.

\bibitem{Ivor} I. Robinson and Y. Ne'eman, {\it Seminar on the embedding problem}, 
Rev. Mod. Phys. {\bf 37} (1965) 201.

\bibitem{Marolf} D. Marolf, {\it Spacetime embedding diagrams for black holes}, Gen. 
Rel. Grav. {\bf 31} (1999) 919.

\bibitem{Deser} S. Deser and O. Levin, {\it Mapping Hawking into Unruh thermal 
properties}, Phys. Rev. {\bf D59} (1999) 064004.

\bibitem{Wang} M.-T. Wang and S.-T. Yau, {\it Isometric embeddings into the Minkowski 
space and new quasi-local mass}, Commun. Math. Phys. {\bf 288} (2009) 919. 

\bibitem{Fronsdal} C. Fronsdal, {\it Completion and embedding of the Schwarzschild 
solution}, Phys. Rev. {\bf 116} (1959) 778.

\bibitem{Paston} S. A. Paston and A. A. Sheykin, {\it Global embedding of the 
Reissner-Nordstr\"om metric in the flat ambient space}, SIGMA {\bf 10} (2014) 003.

\bibitem{Carroll} S. M. Carroll, M. C. Johnson, and L. Randall, {\it Extremal limits 
and black hole entropy}, JHEP {\bf 11} (2009) 109.

\bibitem{Helgi} H. F. R\'unarsson, {\it Limits and special symmetries of extremal black 
holes}, MSc Thesis, Stockholm 2012.

\bibitem{Brigman} G. H. Brigman, {\it Acausal geodesics in the Reissner-Nordstr\o m 
geometry}, Tensor, N. S. {\bf 25} (1972) 267.

\bibitem{Marolf2} J. T. Giblin, D. Marolf, and R. H. Garvey, {\it Spacetime embedding 
diagrams for spherically symmetric black holes}, Gen. Rel. Grav. {\bf 36} (2004) 
83. 

\bibitem{Geroch2} S. K. Chakrabarti, R. Geroch, and C.-b. Liang, {\it Timelike 
curves of limited acceleration in general relativity}, J. Math. Phys. {\bf 24} 
(1983) 597.

\bibitem{CT} W. E. Couch and R. J. Torrence, {\it Conformal invariance under 
spatial inversion of the extreme Reissner-Nordstr\"om black holes}, 
Gen. Rel. Grav. {\bf 16} (1984) 789.

\bibitem{Stefan} S. \AA minneborg, I. Bengtsson, and S. Holst, {\it A spinning 
anti-de Sitter wormhole}, Class. Quant. Grav. {\bf 16} (1999) 363.

\bibitem{Paiva} F. M. Paiva, M. C. Rebou{\c c}as, and M. A. H. MacCallum, {\it Limits of 
spacetimes---a coordinate free approach}, Class. Quant. Grav. {\bf 10} (1993) 1165.

\end{thebibliography}

}
\end{document}